\documentclass[a4paper]{jpconf}
\usepackage{lmodern}
\usepackage{graphicx}
\pdfoutput=1
\usepackage{amsmath}
\usepackage[square,sort&compress,numbers]{natbib}
\providecommand{\abs}[1]{\left\lvert#1\right\rvert}
\begin{document}
\title{Baryon femtoscopy in $\sqrt{s_{\mathrm{NN}}} = 2.76$ TeV Pb--Pb collisions at ALICE}

\author{Jai Salzwedel}

\address{Department of Physics, The Ohio State University, 191 West Woodruff Ave, Columbus, OH 43210, USA }

\ead{jai.salzwedel@cern.ch}

\begin{abstract}
We present femtoscopic results for proton and $\Lambda$ correlation functions measured by ALICE in $\sqrt{s_{\mathrm{NN}}} = 2.76$ TeV Pb--Pb collisions.  Femtoscopic radii are extracted from pp, $\mathrm{\bar{p}\bar{p}}$, and $\mathrm{p\bar{p}}$ pairs.  Comparisons of these radii with those from pion and kaon analyses reveal an approximate transverse mass scaling that is consistent with explanations of radial flow.  Inelastic final state interactions are explored in baryon--antibaryon correlations functions to investigate their relationship with reduced proton yields at LHC energies.
\end{abstract}

\section{Introduction}

The study of two-particle correlations at low relative momentum is commonly known as femtoscopy.  Femtoscopic analyses are capable of measuring spatio-temporal characteristics of heavy-ion collisions, with a particular emphasis made on estimating the homogeneity lengths (also called radii) of the particle-emitting source.  These analyses traditionally study pions \cite{Goldhaber:1960sf,Aamodt:2011mr} because of their ready availability and sensitivity to the spatial scale of the source.  However, measurements of heavier particles such as kaons and baryons can serve to complement the pion results.  One motivation for studying an assortment of heavier particles is to test the hydrodynamic prediction that radial flow should cause the source radii to scale with the transverse mass $m_{\mathrm{T}}$ of the particles \cite{Csorgo:1995bi,Lisa:2005dd}.

Baryon--(anti)baryon correlation functions are also useful in the study of final state interactions (FSI).  Measurements of scattering lengths of pairs such as p$\Lambda$ and $\Lambda\Lambda$ are of interest for rescattering calculations, and, in the latter case, for understanding the properties of neutron stars \cite{SchaffnerBielich:2008kb,Wang:2010gr}.  Baryon-antibaryon correlations allow for the study of annihilation processes.  It has been argued that annihilation in the hadronic rescattering phase should be taken into account when determining particle yields \cite{Werner:2012xh,Karpenko:2012yf,Steinheimer:2012rd}.  This annihilation should result in an anticorrelation in baryon-antibaryon correlation functions.  Results from these femtoscopic analyses may be able to provide insight into the $p/\pi^+$ ratio measured in experiments at the LHC, which falls short of thermal model expectations \cite{Preghenella:2012eu}. 

In this study, we present femtoscopic correlation functions for pp, $\rm{\bar{p}\bar{p}}$, and p$\rm{\bar{p}}$; p$\rm{\bar{\Lambda}}$ and $\rm{\bar{p}\Lambda}$; and $\Lambda\bar{\Lambda}$ pairs.  Source radii have been extracted for pp, $\rm{\bar{p}\bar{p}}$, and p$\rm{\bar{p}}$ pairs, and these are compared with radii from kaon and pion analyses.  

\section{Data analysis}
Femtoscopic analyses have been performed on approximately 40 million Pb--Pb events from ALICE at $\sqrt{s_{\mathrm{NN}}} = 2.76$ TeV.  The Inner Tracking System (ITS), Time Projection Chamber (TPC), and Time-of-Flight detector (TOF) provided tracking for the particles \cite{Aamodt:2008zz}.  Particle identification was performed using the TPC and TOF.  Primary protons were selected based on the distance of closest approach (DCA) to the primary vertex.  $\Lambda$ ($\bar{\Lambda}$) candidates were selected by looking for daughter tracks that match the $\Lambda$'s decay topology.

Two-particle correlation functions were constructed in terms of the one dimensional relative momentum $q_{\mathrm{inv}}=2k^*=\abs{P(m^2_1-m^2_2)/P^2 - q}$, where $P$ and $q$ are the four-vector momentum sum and difference, respectively, and $m_1$ and $m_2$ are the masses of the two particles.

\subsection{Analytical Model}
The correlation functions can be calculated analytically \cite{lednicky82} using 
\begin{equation}
\label{eqLednicky}
C(k^*)= 1+ \displaystyle\sum\limits_{S}\rho_S\left[\frac{1}{2}\abs{\frac{f^S(k^*)}{R}}^2(1-\frac{d_0^S}{2\sqrt{\pi}R})+\frac{2\Re f^S(k^*)}{\sqrt{\pi}R}F_1(2k^*R)-\frac{\Im f^S(k^*)}{R}F_2(2k^*R)\right]
\end{equation}
where $F_1(z) = \int_0^z \! \mathrm{d}x \, e^{x^2-z^2}/z$,  $F_2(z) = (1-e^{-z^2})/z$, and $R$ is the source size. We assume the particles are produced unpolarized, with $\rho_{\mathrm{S}}$ being the fraction of pairs in each total spin state S.  $f^S(k^*)=(1/f_0^S+\frac{1}{2}d_0^Sk^*-ik^*)^{-1}$ is the spin-dependent scattering amplitude, though this analysis measures only spin-averaged values.  Following the techniques of the STAR Collaboration in their $\mathrm{p\Lambda}$ analysis \cite{Adams:2005ws}, the effective radius of interaction $d_0^S$ is also neglected in order to simplify the fit parameters.  The scattering amplitude is dependent upon the complex scattering length $f_0^S$.  The real part of the scattering amplitude can contribute either a positive or a negative correlation, but either way the effect is relatively narrow in $k^*$ (on the order of about one hundred MeV/$c$).  The imaginary part of the scattering amplitude accounts for inelastic processes of baryon-antibaryon annihilation.  Introducing a non-zero imaginary part to the scattering length produces a wide (hundreds of MeV/$c$) negative correlation.  For identical and charged particles, additional terms \cite{Aamodt:2011kd} are necessary to account for quantum interference and the Coulomb interaction, respectively.

\subsection{Experimental correlation functions}
\begin{figure}[h]
\begin{minipage}[b]{18pc}
\includegraphics[width=18pc]{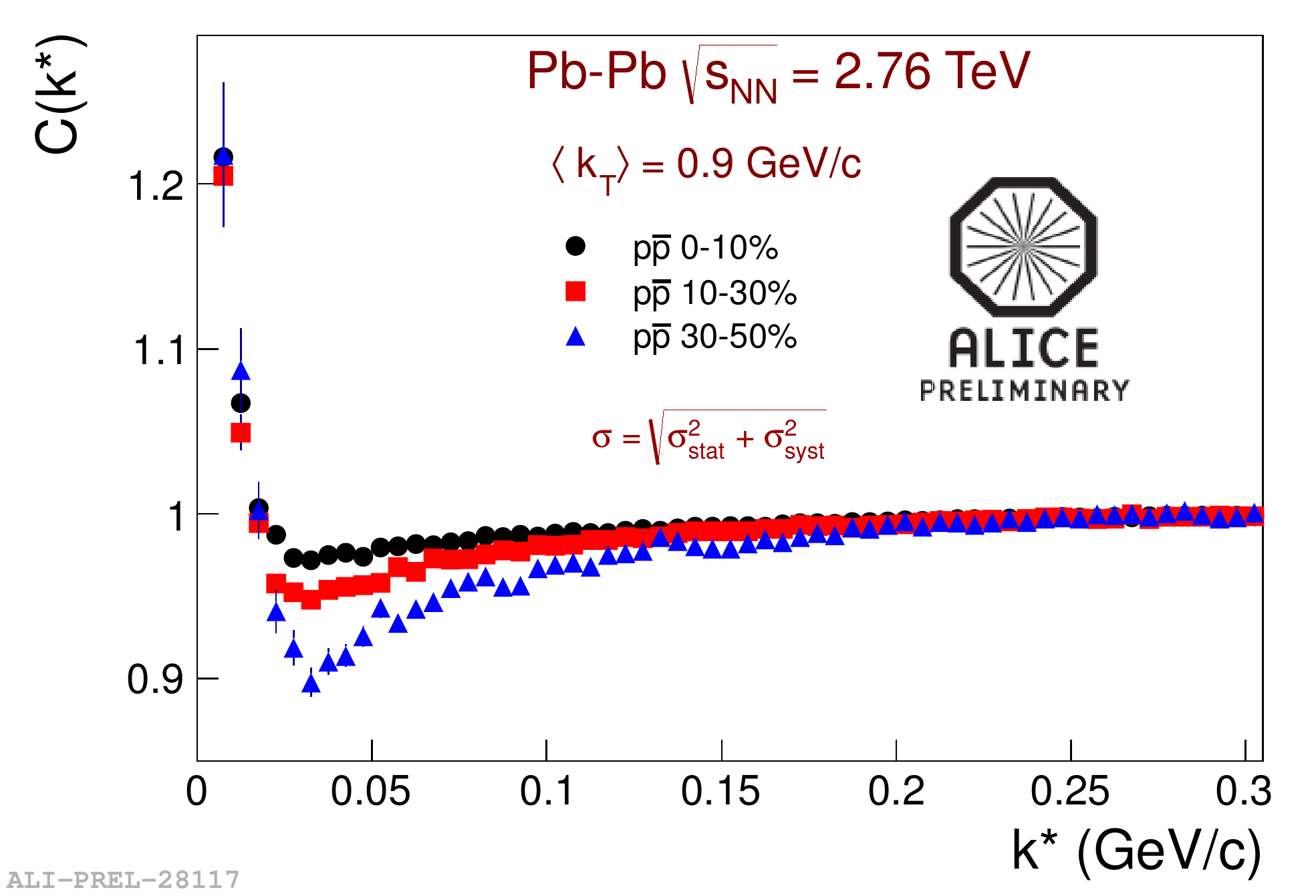}
\caption{\label{fig:pap}Correlation functions for $\mathrm{p\bar{p}}$ in three centrality ranges.}
\end{minipage} 
\end{figure}

\begin{figure}[h]
\begin{minipage}{18pc}
\includegraphics[width=18pc]{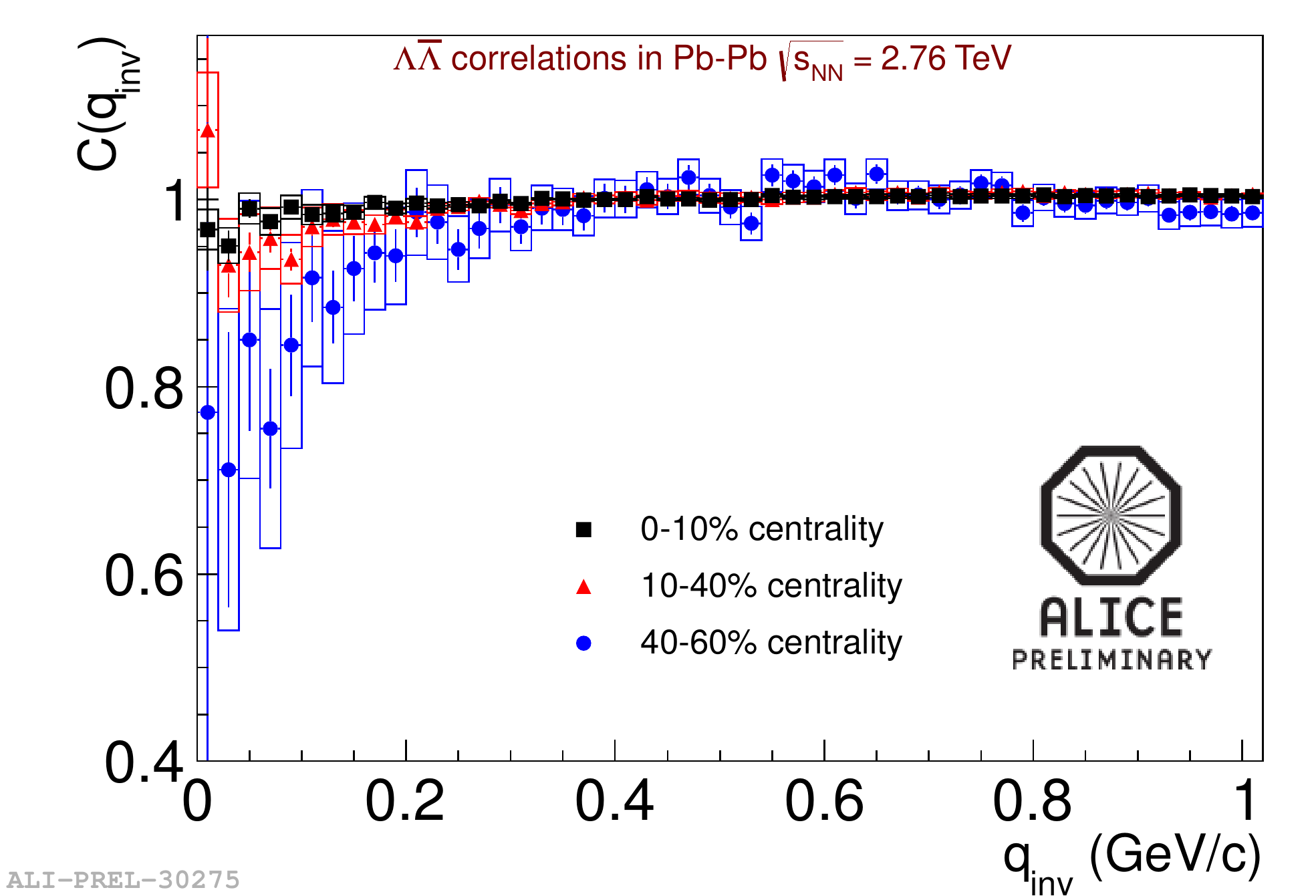}
\end{minipage}\hspace{2pc}
\begin{minipage}{18pc}
\includegraphics[width=18pc]{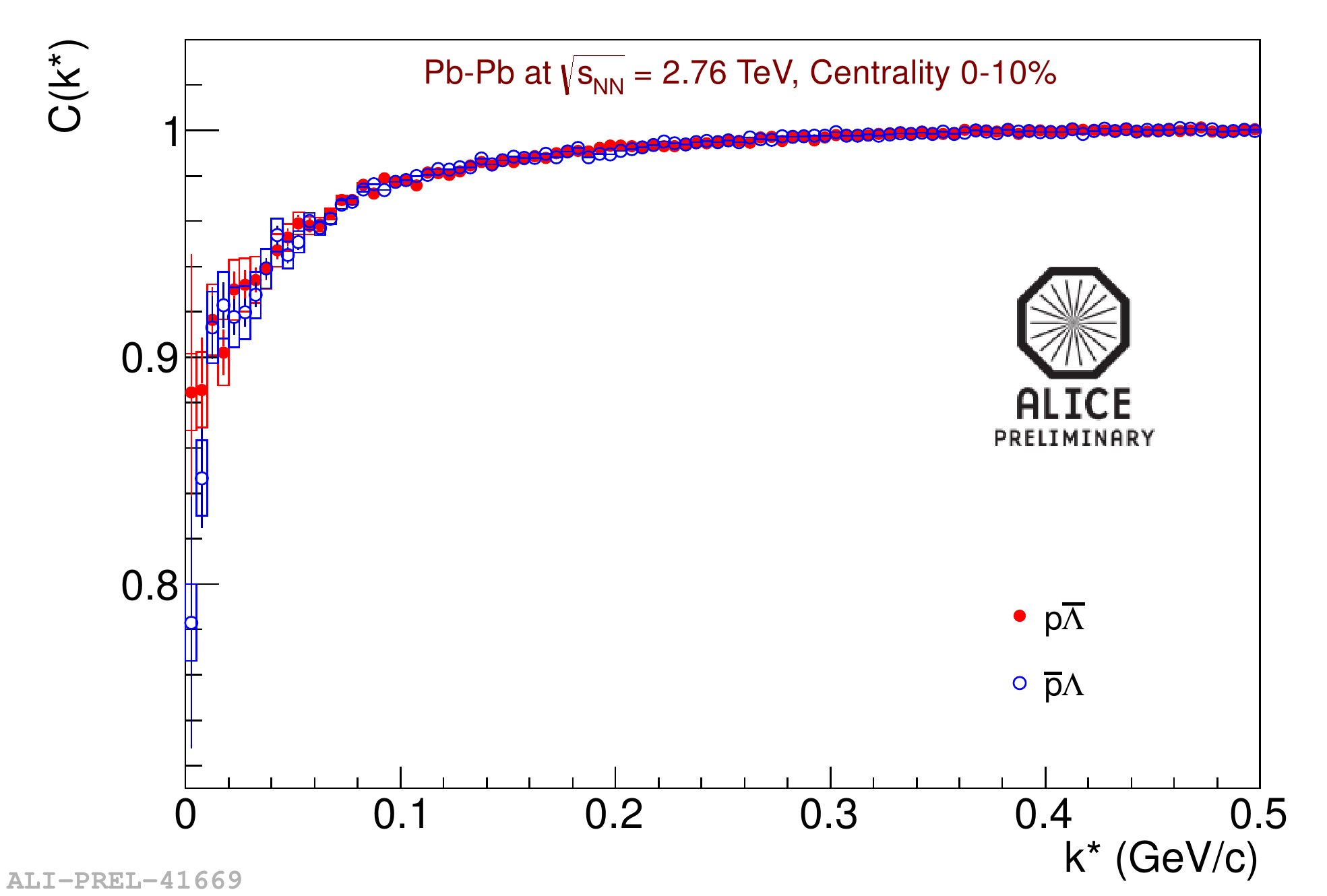}
\end{minipage} 
\caption{\label{fig:pal}Correlation functions for $\mathrm{\Lambda\bar{\Lambda}}$ (left panel) in three centrality ranges.  $\mathrm{p\bar{\Lambda}}$ and $\mathrm{\bar{p}\Lambda}$ (right panel) are shown in the 0-10\% centrality range.}
\end{figure}

Coulomb and strong FSI dominate $\mathrm{p\bar{p}}$ correlations (Fig.~\ref{fig:pap}), while $\mathrm{p\bar{\Lambda}}$, $\mathrm{\bar{p}\Lambda}$, and $\Lambda\bar{\Lambda}$ correlations (Fig.~\ref{fig:pal}) have only strong FSI.  Nonetheless, one common feature is visible in the various correlation functions -- a wide negative correlation in $k^*$.  This shared feature demonstrates that the anticorrelation is not just a result of identical particle-antiparticle annihilation, but an overall expression of inelastic strong FSI resulting from the imaginary part of the scattering amplitude.  

\begin{figure}[h]
\begin{minipage}{18pc}
\includegraphics[width=18pc]{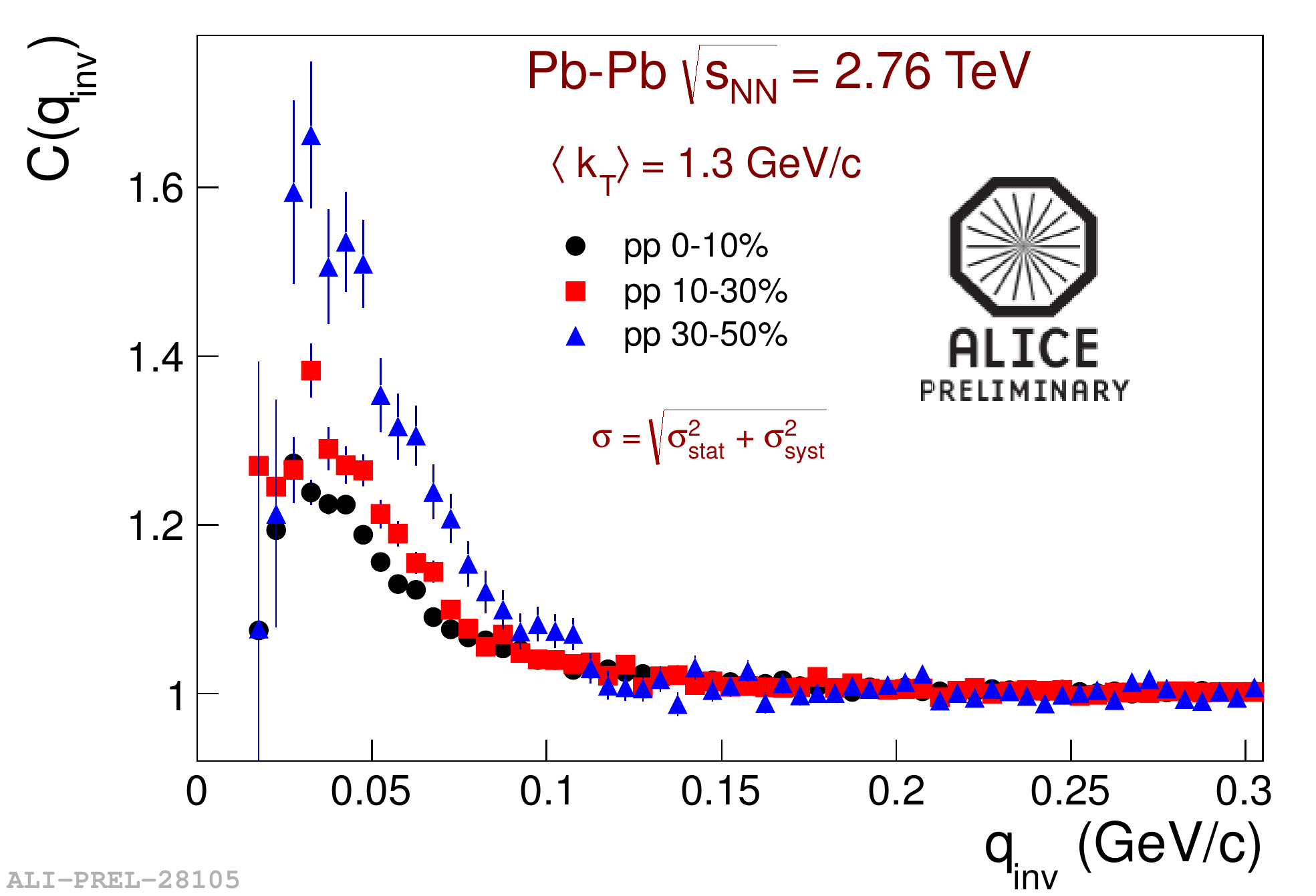}
\end{minipage}\hspace{2pc}
\begin{minipage}{18pc}
\includegraphics[width=18pc]{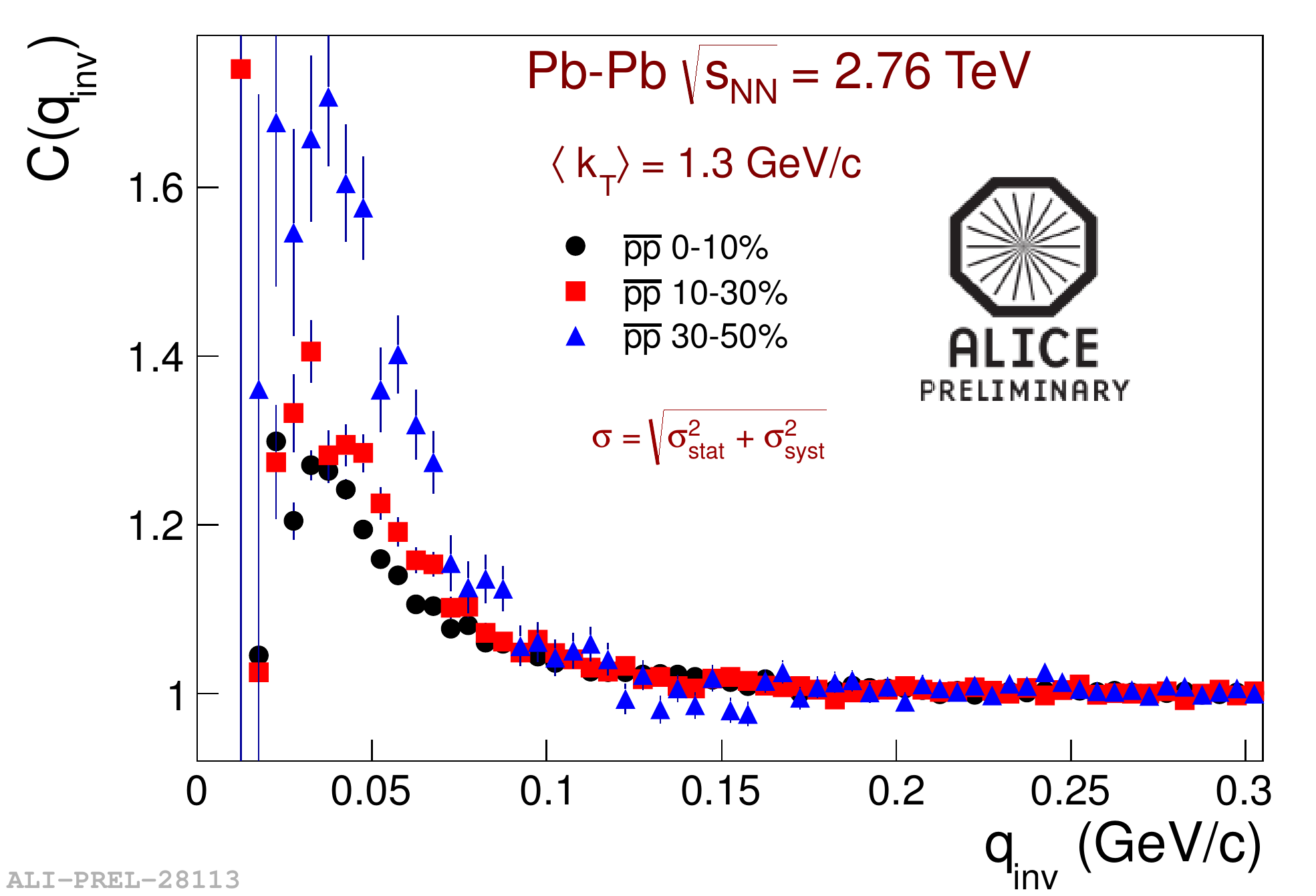}
\end{minipage} 
\caption{\label{fig:pp}Correlation functions for pp (left panel) and $\mathrm{\bar{p}\bar{p}}$ (right panel) in three centrality ranges.}
\end{figure}

Fig.~\ref{fig:pp} shows correlation functions for pp and $\mathrm{\bar{p}\bar{p}}$ pairs in three centrality ranges.  These correlations are affected by Fermi--Dirac statistics, Coulomb interactions, and strong FSI.  The pp and $\mathrm{\bar{p}\bar{p}}$ correlations functions are observed to be consistent with each other within statistical uncertainties, as are the $\mathrm{p\bar{\Lambda}}$ and $\mathrm{\bar{p}\Lambda}$ correlation functions.  For all pairs analyzed, we observe that the strength of the correlation increases as a function of centrality (not shown for $\mathrm{p\bar{\Lambda}}$ and $\mathrm{\bar{p}\Lambda}$).  This behavior is consistent with a decrease of the system size.

Femtoscopic analyses are under way for p$\Lambda$ and $\Lambda\Lambda$, as well as their respective antiparticle pairs.

\subsection{Residual correlations}
In $\mathrm{pp}$ $(\mathrm{\bar{p}\bar{p}})$ correlations, a broad excess is observed in the $0.05$ GeV/$c$ $< k^* < 0.1$ GeV/$c$ range that cannot be explained by correlations coming from the pp wave function.  One potential explanation for this excess is the existence of residual $\mathrm{p\Lambda}$ correlations.  

Not every pp pair will contain two primary protons.  Despite our DCA cuts to select primaries, contamination from secondary particles remains.  For example, in a $\Lambda \rightarrow p + \pi^-$ decay, the proton carries most of the momentum of its parent.  If that $\Lambda$ is primary, its daughter proton will often pass the DCA cuts.  In the case where a signal pair is comprised of one primary proton and one proton from a $\Lambda$ decay, there cannot be any natural proton--proton correlation like quantum interference or FSI between the two particles.  However, the primary proton and the $\Lambda$ may be correlated.  Taking into account momentum smearing due to the decay kinematics, that correlation could also be visible between the primary proton and the $\Lambda$'s daughter proton.  Attempts to measure a pp correlation will therefore see some amount of contamination from these residual $\mathrm{p}\Lambda$ correlations.

We attempt to quantify the residual contamination in our correlation functions by simultaneously fitting the data for both the primary correlation function and the residual correlation function.  For pp correlations with $\mathrm{p\Lambda}$ feeddown, we fit our data using 
\begin{equation}
\label{eq:Residual}
C_{\mathrm{meas}}(k^*_{\mathrm{pp}})= 1 + \lambda_{\mathrm{pp}}[C_{\mathrm{pp}}(k^*_{\mathrm{pp}})-1]+\lambda_{\mathrm{p\Lambda}}[C_{\mathrm{p\Lambda}}(k^*_{\mathrm{pp}})-1],
\end{equation}
where $$C_{\mathrm{p\Lambda}}(k^*_{\mathrm{pp}}) \equiv \displaystyle\sum\limits_{k^*_{\mathrm{p\Lambda}}}C_{\mathrm{p\Lambda}}(k^*_{\mathrm{p\Lambda}})T(k^*_{\mathrm{p\Lambda}},{k^*_\mathrm{pp}}),$$ $C_{\mathrm{p\Lambda}}(k^*_{\mathrm{p\Lambda}})$ is the $\mathrm{p\Lambda}$ correlation function calculated from Eq.~(\ref{eqLednicky}), and $T$ is a THERMINATOR \cite{Chojnacki:2011hb} transform matrix, which generates pairs of particles and determines kinematically how the $k^*$ of the pair transforms when one of the particles decays.

\begin{figure}[h]
\begin{minipage}{18pc}
\includegraphics[width=18pc]{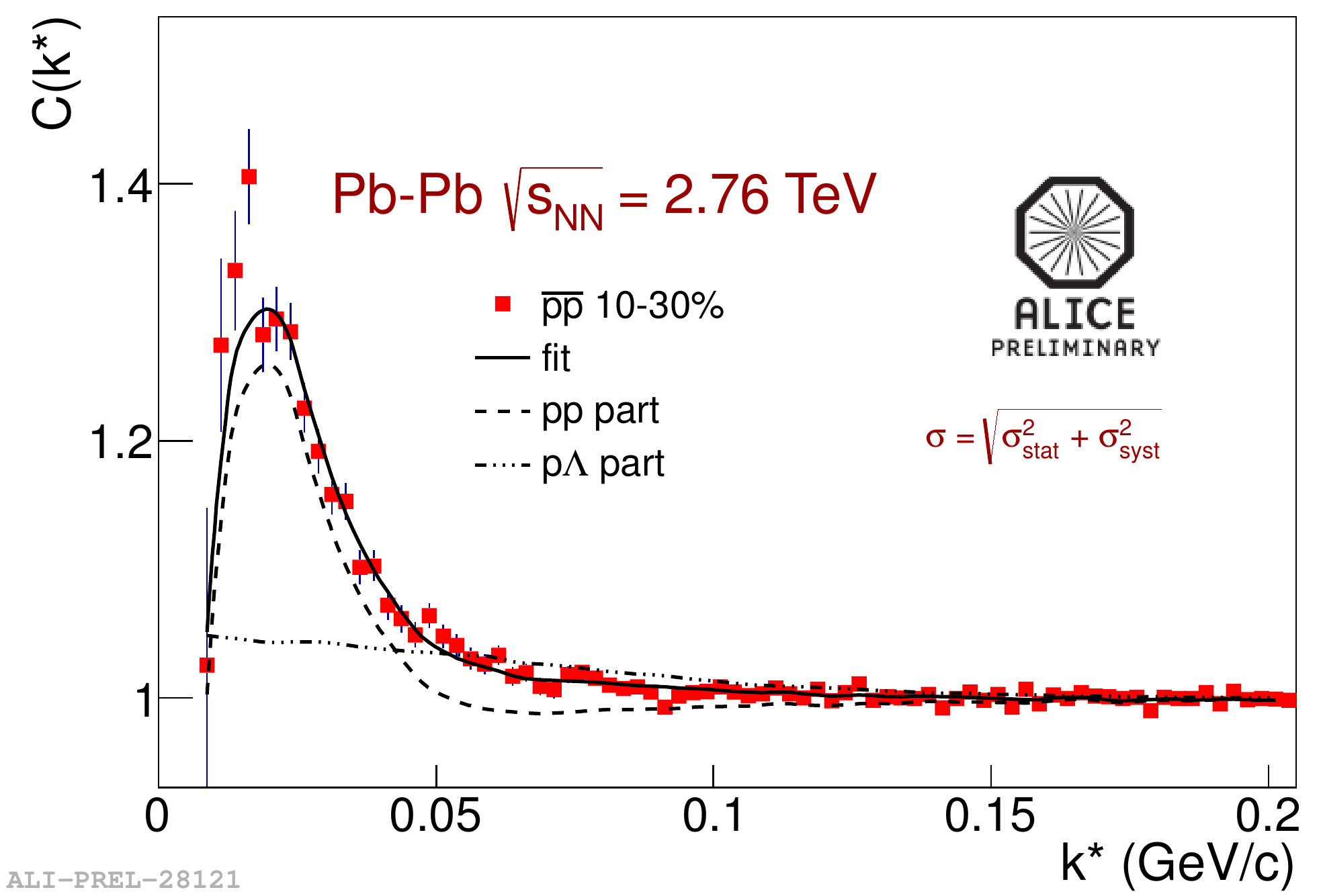}
\end{minipage}\hspace{2pc}
\begin{minipage}{18pc}
\includegraphics[width=18pc]{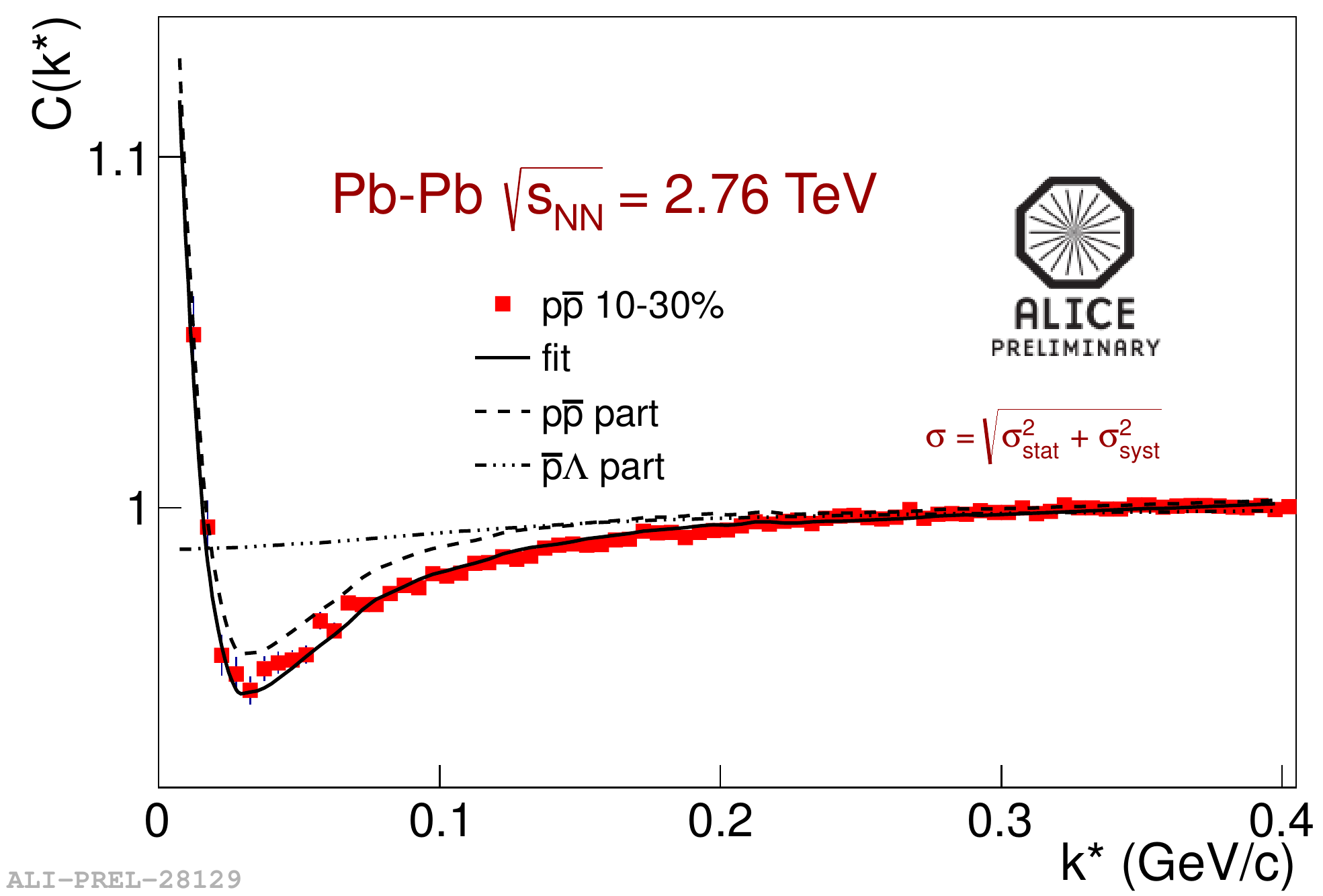}
\end{minipage} 
\caption{\label{fig:apapResiduals}Fits of $\mathrm{\bar{p}\bar{p}}$ (left panel) and $\mathrm{p\bar{p}}$ (right panel) with residual correlation included.  The dashed line shows the primary correlation fit, the dash-dot line shows the residual correlation, and solid line---as the sum of the other two---is the total correlation function.}
\end{figure}

Fig.~\ref{fig:apapResiduals} (left panel) shows ALICE data for a $\mathrm{\bar{p}\bar{p}}$ correlation function that is fit using Eq.~\ref{eq:Residual}.  The figure shows the full fit, as well as the individual $\mathrm{\bar{p}\bar{p}}$ and $\mathrm{\bar{p}\bar{\Lambda}}$ portions.  We see that the combined fit successfully reproduces both the peak at $k^* \approx .03$ GeV/$c$ and the broad excess between $0.03$ and $0.1$ GeV/$c$.  A similar procedure has been successful in fitting $\mathrm{p\bar{p}}$ correlations, as seen in the right panel of Fig.~\ref{fig:apapResiduals}.
 
\section{Fitting Results}

\begin{figure}[h]
\begin{minipage}{18pc}
\includegraphics[width=18pc]{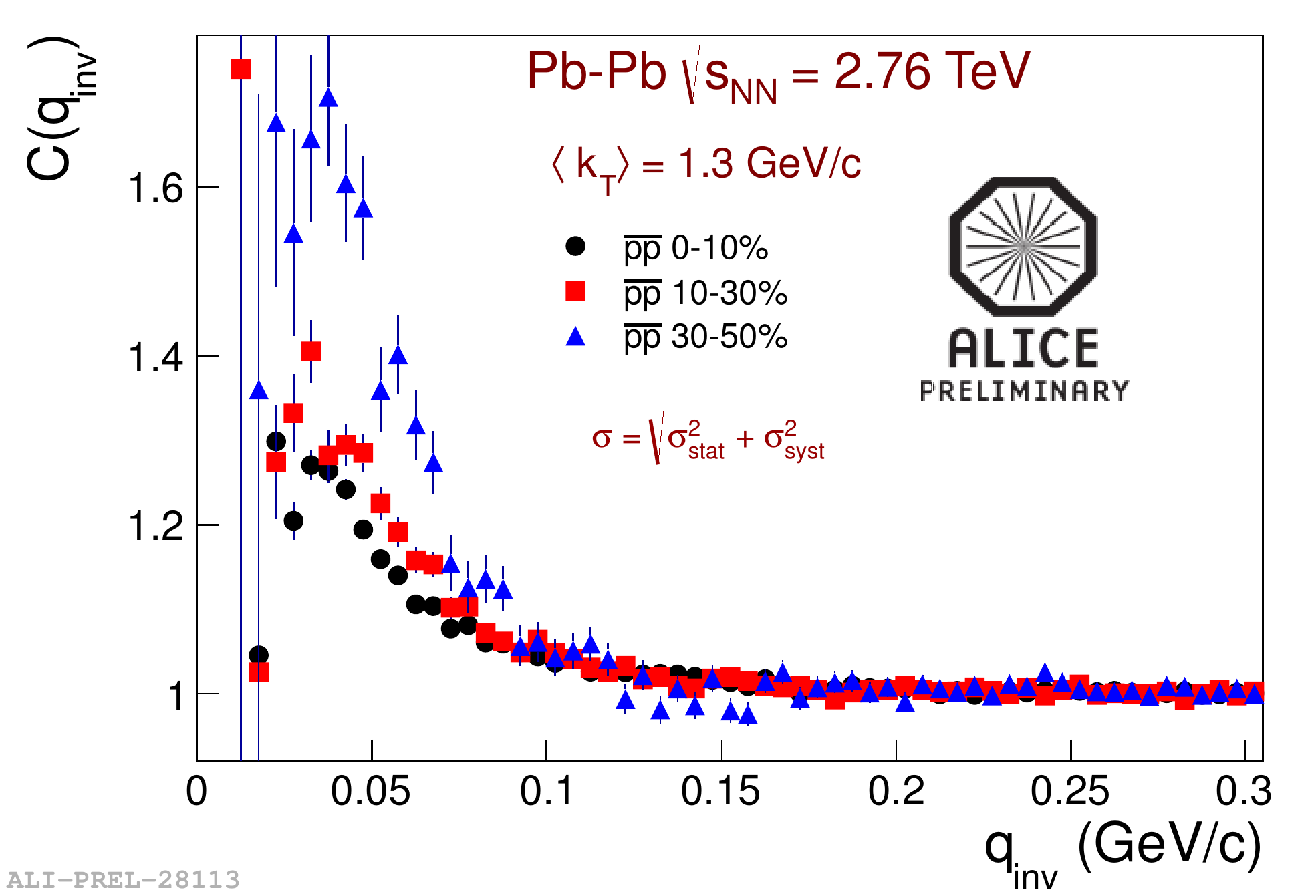}
\caption{\label{fig:ppRadii}Radii extracted from fits to $\mathrm{pp}$, $\mathrm{\bar{p}\bar{p}}$, and $\mathrm{p\bar{p}}$ correlation functions.}
\end{minipage}\hspace{2pc}
\begin{minipage}{18pc}
\includegraphics[width=18pc]{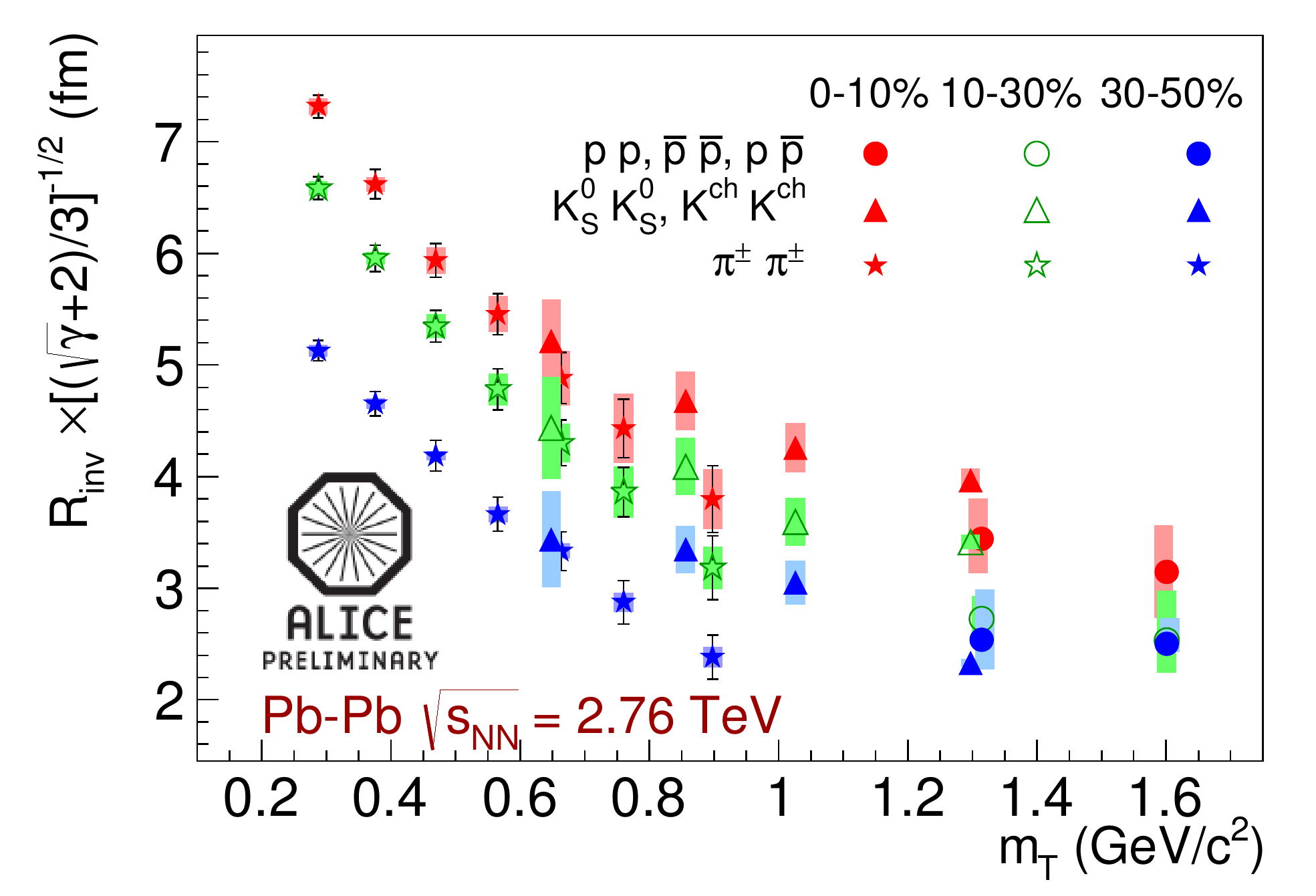}
\caption{\label{fig:mTRadii}Transverse mass dependence of the kinematically scaled radius parameters for pion, kaon, and proton pairs.}
\end{minipage} 
\end{figure}

Preliminary radii have been extracted from $\mathrm{pp}$, $\mathrm{\bar{p}\bar{p}}$, and $\mathrm{p\bar{p}}$ correlation functions using scattering lengths taken from \cite{lednicky82}.  These results are shown in Fig.~\ref{fig:ppRadii} for three centrality bins and two transverse momentum ($k_{\mathrm{T}} = \abs{\vec{p}_{\mathrm{T,1}} + \vec{p}_{\mathrm{T,2}}}/2$) bins.  A tendency is observed for the radii to increase with multiplicity and decrease with $k_\mathrm{T}$.

Fig.~\ref{fig:mTRadii} shows extracted radii as a function of $m_{\mathrm{T}}=(k_{\mathrm{T}}^2 + m_{\mathrm{inv}}^2)^{-1/2}$ for pions, charged and neutral kaons, and protons.  A kinematic scaling factor of $[(\sqrt{\gamma}+2)/3]^{-1/2}$, with $\gamma$ given by $m_{\mathrm{T}}/m_{\mathrm{inv}}$ of the pair, has been appended to the radii to approximate their size in the LCMS frame, where collective flow behavior is expected to be seen.  We observe an approximate $m_{\mathrm{T}}$ scaling that is consistent with accounts of radial flow.

\section{Summary}
Femtoscopic correlation functions have been constructed using protons and $\Lambda$s.  The correlation strength is observed to increase for more peripheral events.  Emission source radii have been extracted from proton correlation functions, and they have been observed to decrease for higher centralities and pair transverse momenta.  An approximate transverse mass scaling of pion, kaon, and proton radii is observed.

\bibliography{iopart-num}

\providecommand{\newblock}{}
\begin{thebibliography}{10}
\expandafter\ifx\csname url\endcsname\relax
  \def\url#1{{\tt #1}}\fi
\expandafter\ifx\csname urlprefix\endcsname\relax\def\urlprefix{URL }\fi
\providecommand{\eprint}[2][]{\url{#2}}

\bibitem{Goldhaber:1960sf}
Goldhaber G, Goldhaber S, Lee W~Y and Pais A 1960 {\em Phys.Rev.\/} {\bf 120}
  300--312

\bibitem{Aamodt:2011mr}
Aamodt K {\em et~al.\/} (ALICE Collaboration) 2011 {\em Phys.Lett.\/} {\bf
  B696} 328--337 (arXiv:\eprint{1012.4035})

\bibitem{Csorgo:1995bi}
Csorgo T and Lorstad B 1996 {\em Phys.Rev.\/} {\bf C54} 1390--1403
  (arXiv:\eprint{hep-ph/9509213})

\bibitem{Lisa:2005dd}
Lisa M~A, Pratt S, Soltz R and Wiedemann U 2005 {\em Ann.Rev.Nucl.Part.Sci.\/}
  {\bf 55} 357--402 (arXiv:\eprint{nucl-ex/0505014})

\bibitem{SchaffnerBielich:2008kb}
Schaffner-Bielich J 2008 {\em Nucl.Phys.\/} {\bf A804} 309--321
  (arXiv:\eprint{0801.3791})

\bibitem{Wang:2010gr}
Wang Y and Shen H 2010 {\em Phys.Rev.\/} {\bf C81} 025801
  (arXiv:\eprint{1002.0204})

\bibitem{Werner:2012xh}
Werner K, Karpenko I, Bleicher M, Pierog T and Porteboeuf-Houssais S 2012 {\em
  Phys.Rev.\/} {\bf C85} 064907 (arXiv:\eprint{1203.5704})

\bibitem{Karpenko:2012yf}
Karpenko I, Sinyukov Y and Werner K 2013 {\em Phys.Rev.\/} {\bf C87} 024914
  (arXiv:\eprint{1204.5351})

\bibitem{Steinheimer:2012rd}
Steinheimer J, Aichelin J and Bleicher M 2013 {\em Phys.Rev.Lett.\/} {\bf 110}
  042501 (arXiv:\eprint{1203.5302})

\bibitem{Preghenella:2012eu}
Preghenella R (ALICE Collaboration) 2012  (arXiv:\eprint{1203.5904})

\bibitem{Aamodt:2008zz}
Aamodt K {\em et~al.\/} (ALICE Collaboration) 2008 {\em JINST\/} {\bf 3} S08002

\bibitem{lednicky82}
Lednicky R and Lyuboshitz V~L 1982 {\em Sov. J. Nucl. Phys.\/} {\bf 35} 770

\bibitem{Adams:2005ws}
Adams J {\em et~al.\/} (STAR Collaboration) 2006 {\em Phys.Rev.\/} {\bf C74}
  064906 (arXiv:\eprint{nucl-ex/0511003})

\bibitem{Aamodt:2011kd}
Aamodt K {\em et~al.\/} (ALICE Collaboration) 2011 {\em Phys.Rev.\/} {\bf D84}
  112004 (arXiv:\eprint{1101.3665})

\bibitem{Chojnacki:2011hb}
Chojnacki M, Kisiel A, Florkowski W and Broniowski W 2012 {\em
  Comput.Phys.Commun.\/} {\bf 183} 746--773 (arXiv:\eprint{1102.0273})

\end{thebibliography}

\end{document}